\documentclass[aps,prl,preprint,groupedaddress]{revtex4-2}
\usepackage{graphicx}

\begin{document}


\def\eqp#1{(\ref{eq:#1})}
\def\eql#1{\label{eq:#1}}
\newcommand{\be}{\begin{equation}}
\newcommand{\ee}{\end{equation}}
\newcommand{\ba}{\begin{eqnarray}}
\newcommand{\ea}{\end{eqnarray}}
\newcommand\ex{\vec{e}_x}
\newcommand\ey{\vec{e}_y}
\newcommand\ez{\vec{e}_z}
\newcommand\p{\partial}
\newcommand\rmH{\mathrm{H}}
\newcommand\rmI{\mathrm{I}}
\newcommand\rmK{\mathrm{K}}
\newcommand\half{\frac{1}{2}}
\newcommand\threehalf{\frac{3}{2}}
\newcommand\quarter{\frac{1}{4}}
\newcommand\Her{\rmH_{-\half}}
\newcommand\Bes{\rmI_{-\quarter}}
\newcommand\hypg{{}_1F_1}
\newcommand\onemnutwo{\frac{2-a}{2}}
\newcommand\nuh{\frac{a-1}{2}}
\newcommand\mnuh{\frac{1-a}{2}}
\newcommand\firstGamma{\Gamma\left(\onemnutwo\right)}
\newcommand\secGamma{\Gamma\left(\mnuh\right)}
\newcommand\Relambda{\mathrm{Re}_\lambda}
\newcommand\vecr{\vec{r}}
\newcommand\vX{\vec{X}}
\newcommand\vu{\vec{u}}
\newcommand\vw{\vec{w}}
\newcommand\bv{\overline{v}}
\newcommand\bp{\overline{p}}
\newcommand\vp{v^\prime}

\title{Viscous vortex layers subject to more general strain and comparison to isotropic turbulence}


\author{Karim Shariff}
\email[]{Karim.Shariff@nasa.gov}
\affiliation{NASA Ames Research Center}

\author{Gerrit E. Elsinga}
\email[]{g.e.elsinga@tudelft.nl}
\affiliation{Laboratory for Aero and Hydrodynamics, Department of Mechanical, Maritime and Materials Engineering, Delft University of Technology, 2628CD Delft, The Netherlands}


\date{\today}

\begin{abstract}
Viscous vortex layers subject to a more general uniform strain are considered.  They include Townsend's steady solution for plane strain (corresponding to a parameter $a = 1$) in which all the strain in the plane of the layer goes toward vorticity stretching, as well as Migdal's recent steady asymmetric solution for axisymmetric strain ($a = 1/2$) in which half of the strain goes into vorticity stretching.  In addition to considering asymmetric, symmetric and antisymmetric steady solutions $\forall a \ge 0$, it is shown that for $a < 1$, i.e., anything less than the Townsend case, the vorticity inherently decays in time: only boundary conditions that maintain a supply of vorticity at one or both ends lead to a non-zero steady state.  
For the super-Townsend case $a > 1$, steady states have a sheath of opposite sign vorticity.  Comparison is made with
homogeneous-isotropic turbulence in which case the average vorticity in the strain eigenframe is layer-like, has wings of opposite vorticity, and the strain configuration is found to be super-Townsend.  
Only zero-integral perturbations of the $a > 1$ steady solutions are stable; otherwise, the solution grows.
Finally, the appendix shows that the average flow in the strain eigenframe is (apart from an extra term) the Reynolds-averaged Navier-Stokes equation.
\end{abstract}


\maketitle


\section{Motivation and Summary of Results}

Vortex configurations subjected to a spatially uniform strain can be used to model local regions of more complicated flows where the strain represents the local potential velocity induced by other vortex structures, typically of larger scale than the region being considered.  
For example,  Burgers' \cite{Burgers_1948} axisymmetrically strained tubular vortex is a good model for the high intensity structures of homogeneous isotropic turbulence \cite{Jimenez_etal_1993}. 
Other examples include
Townsend's Gaussian vortex layer subject to plane-strain \cite{Townsend_1951}, and the celebrated Lundgren spiral which produces Kolmogorov's $k^{-5/3}$ energy spectrum \citep{Lundgren_1982, Gilbert_1993, Pullin_etal_1993, Pullin_etal_1994}.  While we do not consider the instability of strained layers to wavy perturbations, we mention in passing that Townsend's layer is unstable to the formation of concentrated tubular structures \citep{Lin_and_Corcos_1984, Passot_etal_1995}.  This suggests a similar fate for other solutions presented below.

Recently, Migdal \cite{Migdal_2021} presented a steady asymmetric vortex layer solution for the case of axisymmetric strain ($a = 1/2$ below).  In this solution the vorticity decays algebraically on one side of the layer and as a Gaussian on the other.  It was the desire to interpret this solution that led to the present note.  We conclude that in this configuration, stretching cannot keep up with diffusion unless there is a supply of vorticity from the algebraically decaying side.  The symmetric solution, which Migdal did not consider, corresponds to algebraic decay on both sides, while the antisymmetric solution corresponds to annihilation of vorticities of opposite sign.  The above conclusions apply equally well for values of $0 < a < 1$, though less and less of a supply of vorticity is needed as $a \to 1$ (for the symmetric and asymmetric steady solutions).   For $a = 1$ we have Townsend's plane strain case in which a steady state is reached with zero boundary conditions on both sides and non-zero integrated vorticity in the initial condition.  For $a > 1$, steady states exist with inflow of opposite sign vorticity at one or both ends.  Only zero integral perturbations of these states relax back to the steady state; otherwise, they grow.

It is known that in homogeneous-isotropic turbulence, vorticity tends to align with the direction of the intermediate strain rate \cite{Ashurst_etal_1987}.  However, when the contribution of the strain induced by the local vorticity is removed, it is found that the vorticity is aligned with the direction of the largest background strain \cite{Hamlington_etal_2008}.  In our set-up, this corresponds to the super-Townsend case $a > 1$.  Elsinga etal.~\cite{Elsinga_etal_2017} studied the averaged local vorticity structure of homogeneous isotropic turbulence simulations in the strain eigenframe.  The structure consists of a vortex layer and two tube-like vortices adjacent to it.
Interestingly, the vorticity component in the direction of the largest principal (background) strain versus the direction normal to the layer is symmetric and changes sign.  This is the type of steady solution we obtain for the super-Townsend cases.  The background strain in the turbulence simulations is also super-Townsend.  However, the negative vorticity wings in the turbulent case have a higher amplitude and are more extended (Figure \ref{fig:Elsinga}d below) than in our steady solution.  Finally, the appendix shows that the averaged local flow in the strain eigenframe is governed (apart from one term) by the Reynolds averaged Navier-Stokes equation.  This may help to further understand its structure.

\section{Analysis}

We consider the unidirectional shear flow and associated vorticity
\be
   u_x = U(z, t), \hskip 0.5truecm  \omega_y(z, t) = U'(z, t) \equiv G(z, t), \eql{shear}
\ee
subjected to an irrotational strain written in principal coordinates as
\be
\vec{u}_\mathrm{strain} = \alpha \left( (1-a) x \ex + a y \ey - z\ez\right). \eql{strain}
\ee
Equations \eqp{shear} and \eqp{strain} assume that the vorticity is aligned with one of the principal axes of strain.
A more general set up would allow the vorticity to be arbitrarily oriented with respect to the strain axes; in this case the
vorticity would undergo a period of alignment.

The strain coefficients add up to zero to respect incompressibility.
We choose $\alpha, a > 0$ since to counteract diffusion we want a compression in $z$ and stretching along $y$.
For $a = 0$ all of the straining in the plane ($xy$) of the layer goes into advection and none into vortex stretching.
For $a = 1/2$ we get Migdal's case of axisymmetric strain in which half of the straining goes into vorticity stretching and half into advection.
For $a = 1$ we recover Townsend's case of strain in the $yz$ plane (independent of $x$).  In this case, \textit{all} of the straining flow in the plane of the layer goes into vorticity stretching.  
We mention in passing that the Townsend case is amenable to conformal mapping in the $yz$ plane for generating steady solutions for non-uniform strains \cite{Bazant_and_Moffatt_2005}. 
The case $a > 1$ corresponds to even greater $y$ stretching than Townsend's case and has compression along $x$; we refer to it as being ``super-Townsend.''

The only non-trivial component of the vorticity equation is the $y$-component and it gives the linear PDE:
\be
   \p_t G - \alpha z \p_z G = a\alpha G + \nu \p_{zz} G, \eql{ve}
\ee
where $\nu$ is the kinematic viscosity.
The second term on the left side of \eqp{ve} represents advection while the first term on the right side represents stretching.  Setting the time derivative equal to zero gives the ODE:
\be
   \nu G''(z) + \alpha z G'(z) + a \alpha G(z) = 0, \eql{ode}
\ee
whose general solution given by Mathematica is 
\be
   G(z, a) = c_1 G_1(z, a) + c_2 G_2(z, a).
\ee
If we define
\be
   \eta = \frac{z}{\sqrt{2}\delta} \mathrm{\ with\ } \delta \equiv (\nu/\alpha)^{1/2}
\ee
as a measure of the sheet thickness, then
\ba
G_1(\eta, a) &=& \exp\left(-\eta^2\right) H_{a-1}\left(\eta\right), \\
G_2(\eta, a) &=& \exp\left(-\eta^2\right) \hypg\left( \frac{1-a}{2}, \half, \eta^2\right).
\ea
The function $\hypg$ is Kummer's confluent hypergeometric  function and
$H_{a-1}$ is a Hermite function defined as
\be
H_{a-1}(\eta) = 2^{a-1} \sqrt{\pi}
\left[
\frac{1}{\firstGamma}\hypg\left(\mnuh,\half,\eta^2\right) -
\frac{2\eta}{\secGamma}\hypg\left(\onemnutwo, \threehalf, \eta^2\right)
\right]. \eql{Hermite}
\ee%
The first term in $G_1$ (from the first term in eq. \eqp{Hermite}) is proportional to $G_2(z; a)$, i.e., it is not linearly independent and we may discard it.  However, since $G_1(z, a)$ as it stands is the solution (for $a = 1/2$) presented by Migdal \cite{Migdal_2021}, we shall not alter it.
\begin{figure}
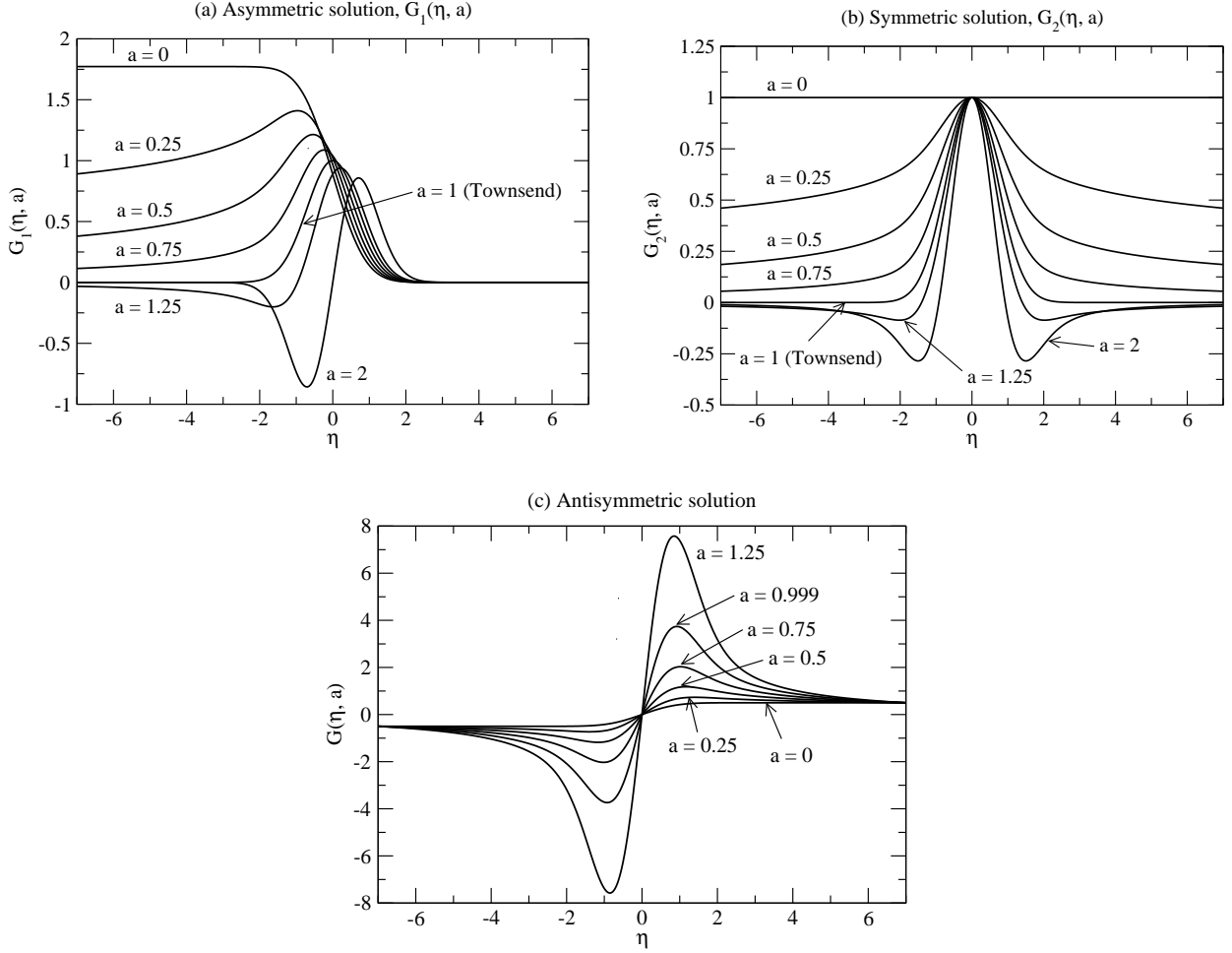

\centering
\includegraphics[width=3.1truein]{new_fig1a.eps}
\hfill
\includegraphics[width=3.1truein]{new_fig1b.eps}
\vskip 0.5truecm
\centering
\includegraphics[width=3.1truein]{new_fig1c.eps}
\caption{(a) and (b) Two linearly independent solutions, $G_1(\eta; a)$ and $G_2(\eta; a)$ to the ODE \eqp{ode} plotted for various strain configurations $a$. (c) Antisymmetric solutions constructed from a suitable linear combination of $G_1(\eta)$ and $G_2(\eta)$.}
\label{fig:G1G2}
\end{figure}

Figure~\ref{fig:G1G2} plots $G_1(\eta, a)$ and $G_2(\eta, a)$ for various strain configurations.  %
The two functions are asymmetric and symmetric, respectively.  The asymmetry of $G_1(\eta, a)$ comes from the $\eta$ in the second term of the Hermite function \eqp{Hermite}.  Note that due to symmetry of the ODE \eqp{ode} under $z \to -z$, a plus sign for the second term instead of the minus sign should be an equally valid solution, i.e, the mirror image of $G_1(\eta, a)$ should be an equally valid solution.  While this can be achieved with a suitable choice of $c_1$ and $c_2$, the $z \to -z$ symmetry is made more obvious if we use
the pair of solutions
\be
 G^{\pm} \equiv \left[
\frac{1}{\firstGamma}\hypg\left(\mnuh,\half, \eta^2\right) \pm
\frac{2\eta}{\secGamma}\hypg\left(\onemnutwo, \threehalf, \eta^2\right)
\right], \eql{pm}
\ee
which are mirror images of each other.  The functions $G^+ \pm G^{-}$ which are symmetric and antisymmetric, respectively, can also be used as a basis.

Consider a finite domain $\eta \in [-L, L]$.
The case $a = 0$ corresponds to plane strain perpendicular to the vorticity so stretching is absent.  To achieve a steady state in this case we need a source of uniform vorticity at one or both ends; for example, a vortex patch (see the $a = 0$ case in Figure~\ref{fig:G1G2}a).
As $a$ increases one requires smaller boundary values until for $a = 1$ (Townsend), the required value is exponentially small in $L^2$.

For the super-Townsend case ($a > 1$), refer to the curves for $a = 1.25$ or $a = 2$  in Figure~\ref{fig:G1G2}.
The symmetric and asymmetric steady solutions show opposite sign vorticity entering one or both boundaries and changing sign before reaching $\eta = 0$.  Since neither advection, stretching, nor dissipation can change the sign of the vorticity, to achieve such a steady state requires special initial conditions; this will be discussed in more detail in the next section.

The functions $G_1(\eta)$ and $G_2(\eta)$ may be combined to form antisymmetric solutions; see  Figure~\ref{fig:G1G2}c.  One can do this for all $a$ except $a = 1$ because both $G_1(\eta, 1)$ and $G_2(\eta, 1)$ are symmetric: $G_1(\eta, 1) \propto G_2(\eta, 1) = \exp(-\eta^2)$.  This is why in Figure~\ref{fig:G1G2}c the solution for $a = 0.999$ is plotted instead.  However, since one can let $a \to 1$ arbitrarily closely, an antisymmetric solution does exist in the limit.  To confirm this, the unsteady code was run for $a = 1$, an initial condition of zero, and antisymmetric boundary conditions $G(-7) = -0.5$ and $G(7) = 0.5$.  The run converged to a steady solution as $t \to \infty$.

\section{Connecting to turbulence}

\begin{figure}
\centering
\includegraphics[width=3.1truein]{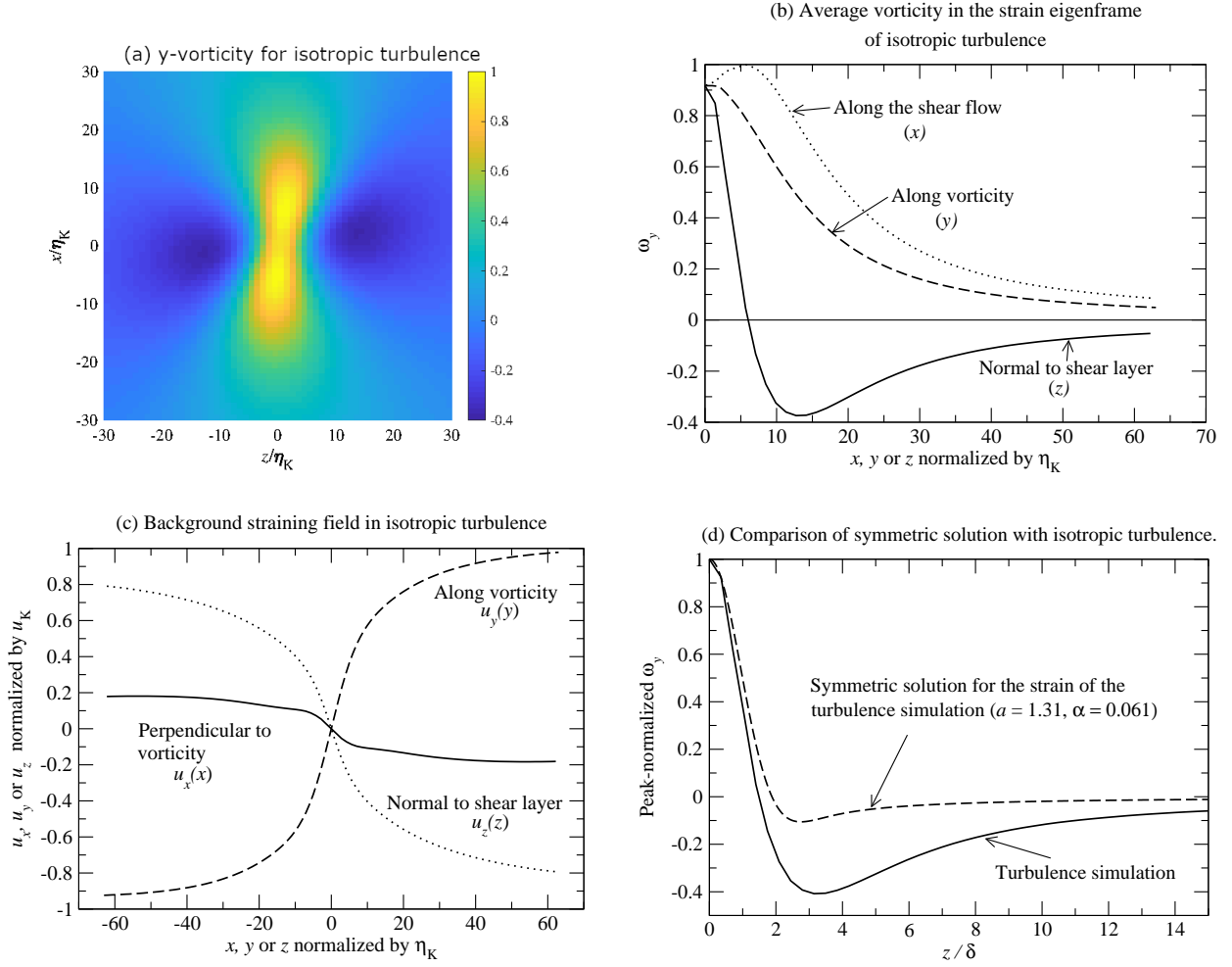}
\hfill
\includegraphics[width=3.1truein]{new_fig2b.eps}
\vskip 0.5truecm\centering
\includegraphics[width=3.1truein]{new_fig2c.eps}
\hfill
\includegraphics[width=3.1truein]{new_fig2d.eps}
\caption{From the eigenframe analysis of a direct simulation of isotropic turbulence \cite{Elsinga_etal_2017} showing the average local flow in the strain eigenframe.  Taylor microscale Reynolds number $\Relambda = 433$.  (a) Contours of vorticity $\omega_y$ in the $xz$ plane.  (b) Profiles of $\omega_y$ along three axes.  (c) Profiles of the background velocity along the three axes.  (d) Comparison of the simulation $\omega_y(z)$ against the symmetric solution $G_2(z/ (\sqrt{2}\delta), a)$ for the strain configuration in the simulation ($\alpha = 0.061, a = 1.31$).}
\label{fig:Elsinga}
\end{figure}

Elsinga etal.~\cite{Elsinga_etal_2017} studied the average local flow for homogeneous isotropic turbulence simulations in the strain eigenframe.  They found that the vorticity consists of a shear layer-like structure; see Figure~\ref{fig:Elsinga}a. 
The vorticity component plotted is $\omega_y$; it is the dominant vorticity component in the $xz$ plane, the others being at most $0.4\%$. 
This figure is oriented such that the shear velocity $u_x = \mathrm{constant} = 0$ along $z = 0$ which we call the shear-layer centerline.
A streamline plot, for which we refer the reader to \cite{Elsinga_etal_2017}, reveals that at each end of the layer,  there is an converging spiral indicating a strained tube-like structure.
The entire structure has (full-width-half-maximum) dimensions of $(39.3 (x) \times 27.4 (y) \times 6.8 (z))\eta_\rmK$, where $\eta_\rmK$ is the Kolmogorov scale.  
The interesting feature is the presence of a sheath of negative vorticity similar to the symmetric solution for the super-Townsend cases.
Figure~\ref{fig:Elsinga}b shows profiles of $\omega_y$ along the three axes; they are symmetric and only one half is plotted.  
The negative sheath can be observed in the profile (solid line) normal to the layer ($z$). 
One therefore wonders whether the background strain is super-Townsend in the simulation.  
Figure~\ref{fig:Elsinga}c plots velocities (in Kolmogorov units) for the background straining flow obtained by removing the contribution to the strain from the local vorticity using Biot-Savart integration \cite{Elsinga_etal_2017}.  The fact that $u_x(x)$ for the background flow has a negative slope means that $1 - a < 0$ so that indeed $a > 1$.  Evaluating slopes at the origin gives $a = 1.31$ and $\alpha = .061$ (in Kolmogorov units).  Note that in Kolmogorov units $\epsilon = \nu = 1$ where $\epsilon$ is the dissipation rate.  Therefore in Kolmogorov units $\delta \equiv (\nu/\alpha)^{1/2} = (1/0.061)^{1/2} = 4.05$ for the simulation.  
Figure~\ref{fig:Elsinga}d compares the simulation profile with the symmetric solution $G_2(z/(\sqrt{2}\delta), a)$ for $a = 1.31$ and $\delta = 4.05$.  Clearly, in the turbulent simulation the negative sheath has a greater amplitude and range.  In fact, the negative area under the curve is larger than the positive.  The results of the next section then imply that if the boundary condition and strain were fixed, the negative vorticity would grow without bound.  

It would be of interest to investigate the origin of the negative vorticity in the simulation by examining individual fields in the sample.  
The Appendix shows that the equation governing the average flow in the strain eigenframe is (apart from one term) just the Reynolds averaged Navier-Stokes equation.  Hence, another approach for understanding the structure of the average flow would be to obtain terms in this equation from the simulation. The present work leads one to expect that divergence of the Reynolds stress and extra term will be sub-dominant near the origin.

The asymmetric solutions $G_1(z)$ are reminiscent of the measured vorticity field at the turbulent-nonturbulent interface of many flows, conditionally averaged with respect to the interface; see Figs.~5 and 6 in the review article \cite{da_Silva_etal_2014} and Fig.~1d in \cite{Westerweel_etal_2005}.  These interfaces are also strained \cite{Elsinga_and_Da_Silva_2019}.  The interfaces of large vorticity voids in isotropic turbulence might be similar.  The asymmetric solutions are also reminiscent of the edge layer at the boundary of a laminar vortex ring \cite{Shariff_and_Krueger_2018}.  Vorticity diffusing across the dividing streamline is subject to a strain induced by the fact that the ring is moving as a whole.  The edge layer imposes a Robin-type boundary condition on the vorticity, which is similar to Newton's law of convective cooling at the boundary of a conducting solid.

Note that the structure in Figure~\ref{fig:Elsinga}a has a slight tilt in orientation relative to shear-layer centerline defined earlier: the positive vorticity is tilted clockwise while the negative vorticity is tilted counter-clockwise.  This implies an asymmetry in $\omega_y(z)$ profiles when plotted at $x \neq 0$.  Hence the tilt could be due to asymmetric vorticity in the background flow.

In the discussion section of his paper, Townsend \cite{Townsend_1951} argues that the averaged product $\varpi$ of principal strains in isotropic turbulence must be negative because it is proportional to the derivative skewness which is known to be negative.  For our set-up $\varpi = \gamma a (a-1)$ and since $\gamma, a > 0$, we need $a < 0$ for $\varpi < 0$.  That is, the strain must be sub-Townsend according to this reasoning.  This is contrary to the previous paragraph and the resolution likely lies in the fact that the derivative skewness also includes contributions from the local vorticity field.  Hence, Townsend's argument should not be used to infer the configuration of background strain.

\section{Time-dependent behavior}

The unsteady linear PDE \eqp{ve} was solved numerically.  Since there are dimensions of length and time, we are free to set $\alpha = 2\nu = 1$ which makes $z = \eta$.  The remaining 
parameters are the strain configuration $a$ and those that involve the initial and/or boundary values of the vorticity.  Denoting the peak magnitude of the vorticity at any instant as $|\omega|_\mathrm{max}(t)$, a time-dependent Reynolds number may be defined as
\be
   \mathrm{Re} \equiv \frac{|\omega|_\mathrm{max}(t)| \delta^2}{\nu} = |\omega|_\mathrm{max}(t),
\ee
for our choice of units.  Since the problem is linear and the vorticity amplitude does not matter, there is no Reynolds number dependence.
The domain is $\eta \in [-L, L]$ with $L = 7$ and various Dirichlet boundary conditions are applied.
\begin{figure}
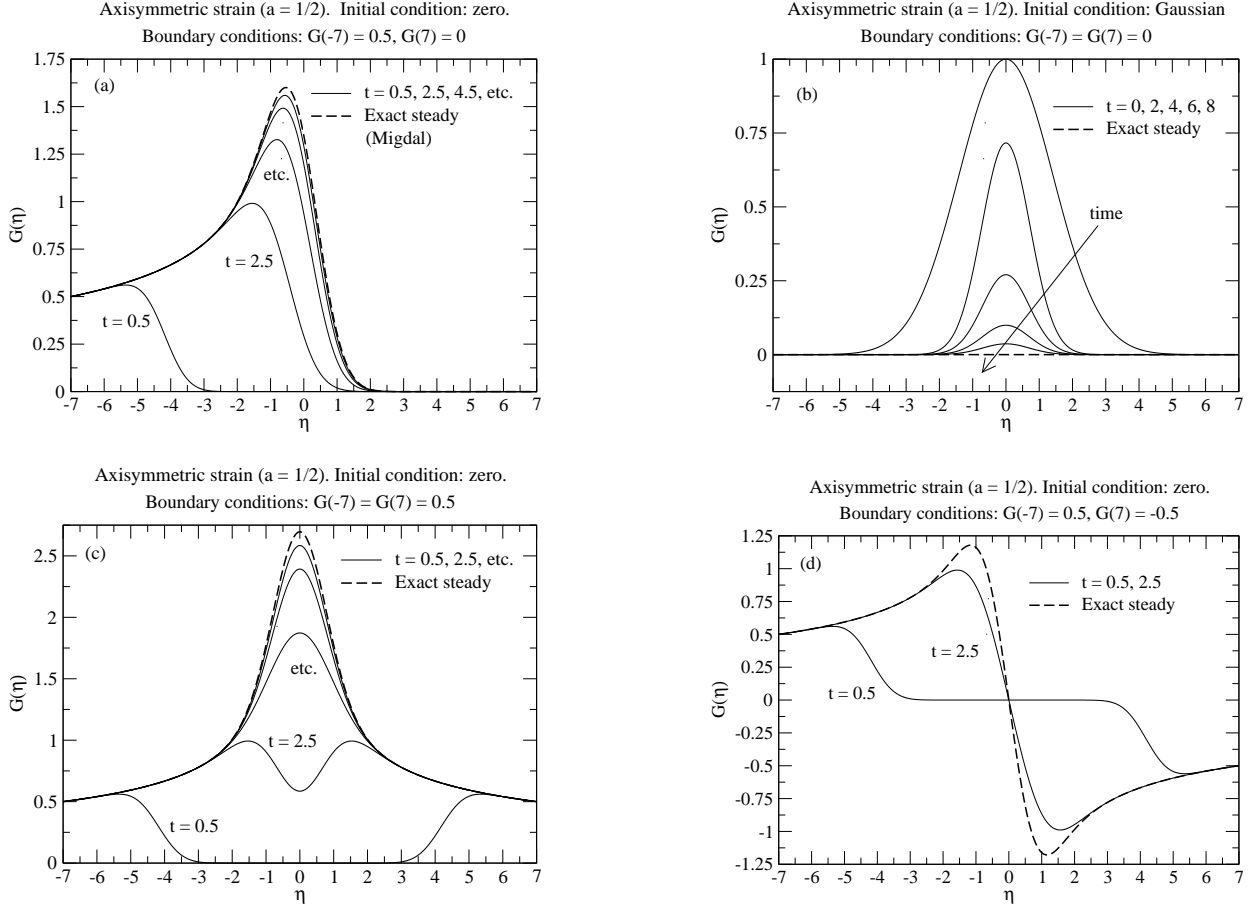

\includegraphics[width=2.8truein]{new_fig3a.eps} \hfill
\includegraphics[width=2.8truein]{new_fig3b.eps}
\vskip 0.4 truecm
\centering
\includegraphics[width=2.8truein]{new_fig3c.eps}
\hfill
\includegraphics[width=2.8truein]{new_fig3d.eps}
\caption{Four axisymmetric strain cases ($a = 1/2$).  (a) Relaxation to Migdal's \cite{Migdal_2021} asymmetric steady state with supply of vorticity at the left end.  (b) Decay to zero with no supply at either end.  (c) Relaxation to a symmetric solution with symmetric supply at both ends.  (d) Relaxation to antisymmetric steady-state with supply of positive and negative vorticity followed by annihilation.}
\label{fig:two}
\end{figure}

Figure~\ref{fig:two} is for axisymmetric strain ($a = 1/2$) and panel (a) shows relaxation to the exact asymmetric steady state of Migdal \cite{Migdal_2021} with an initial condition of zero and asymmetric boundary conditions $G(-7) = 0.5$ and $G(7) = 0$.  Keeping in mind that there is advection of vorticity toward the origin, we see that, to maintain Migdal's asymmetric solution, one needs a continual supply of vorticity at one end; otherwise, the solution would eventually decay to zero.
Figure~\ref{fig:two}b shows that the solution decays if conditions of zero are applied at both ends.  Panel (c) shows relaxation to the symmetric solution with a symmetric supply of vorticity at both ends and an initial condition of zero.  Figure~\ref{fig:two}d shows relaxation to the antisymmetric solution in which vorticity of opposite sign enters the domain, is amplified by stretching and then annihilates.

\begin{figure}
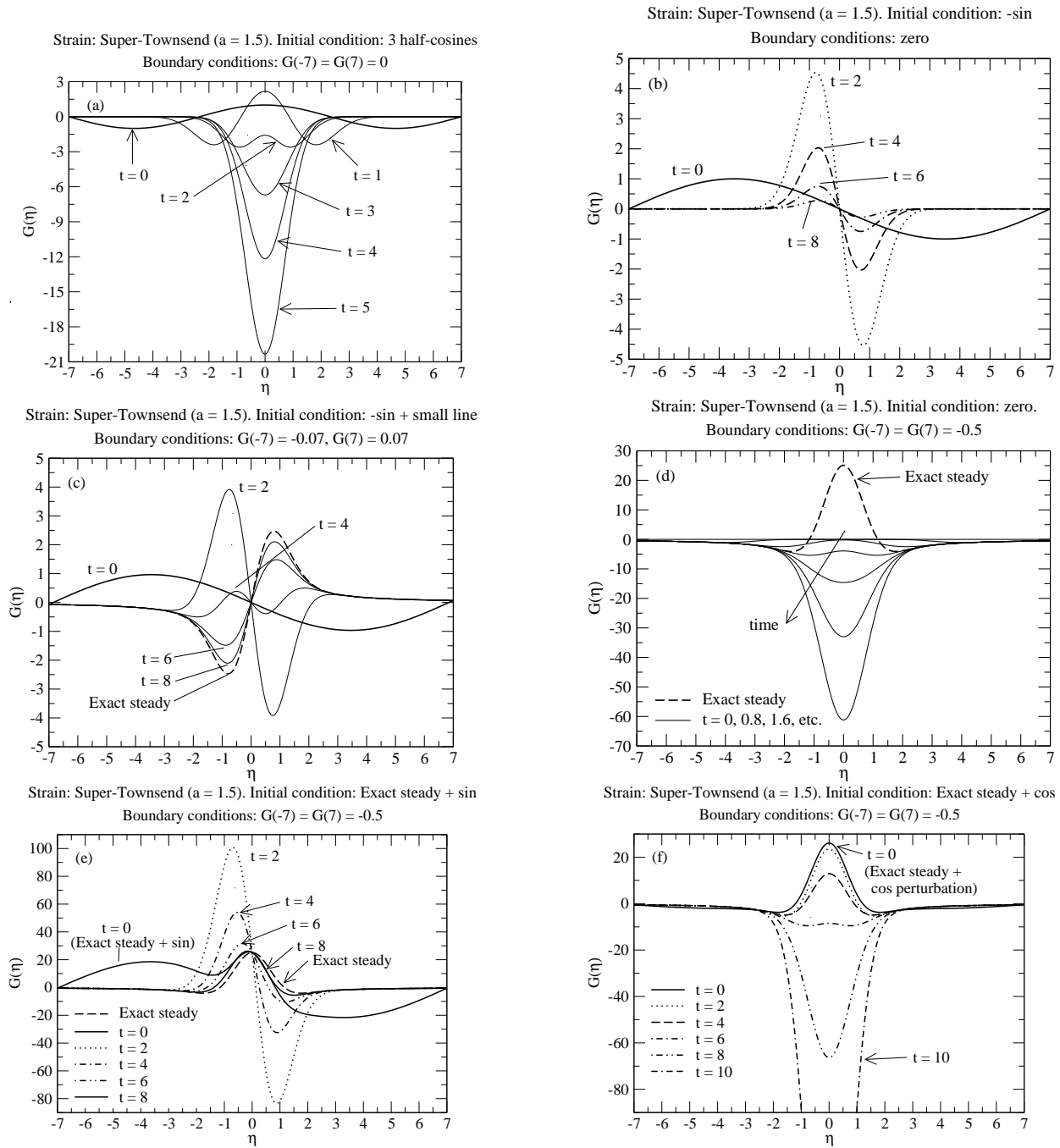

\includegraphics[width=2.9truein]{new_fig4a.eps}\hfill
\includegraphics[width=2.9truein]{new_fig4b.eps}
\includegraphics[width=2.9truein]{new_fig4c.eps} \hfill
\includegraphics[width=2.9truein]{new_fig4d.eps}
\includegraphics[width=2.9truein]{new_fig4e.eps} \hfill
\includegraphics[width=2.9truein]{new_fig4f.eps}
\caption{Super-Townsend cases $a = 1.5$: 
(a) Exponential growth of negative vorticity with three half-cosine initial condition and zero boundary conditions.  
(b) Decay with a sine wave initial condition and zero boundary conditions.  
(c) Adding a small line to the sine wave leads to relaxation to the steady solution.
(d) Failure to relax to the steady solution.
(e) Relaxation back to the steady state after a conservative perturbation.
(f) Failure to relax back to the steady state after a non-conservative perturbation.}
\label{fig:st}
\end{figure}

Finally, we turn attention to some super-Townsend cases (Figure~\ref{fig:st}, $a = 1.5$).  We begin with homogeneous boundary conditions $G(-7) = G(7) = 0$.  For this case there is only the trivial steady solution $c_1 = c_2 = 0$ and so we expect that an unsteady solution will either decay or grow when starting from an arbitrary initial condition.  Panel (a) shows that with an initial condition consisting of three half-cosines, the vorticity grows exponentially as a near-Gaussian.  It grows in the negative direction because the initial condition has a negative integral.
The corresponding Townsend case ($a = 1$) would saturate to a near-Gaussian steady state.
On the other hand, the sine wave initial condition,
\be
   G(\eta, 0) = -\sin (2\pi\eta/\lambda), \hskip 0.5truecm \lambda = 2L, \eql{sin}
\ee
which has zero integral, decays to zero (Figure \ref{fig:st}b).  Next, we make a slight change by adding a small linear function, $0.01\eta$, to 
\eqp{sin} and enforce the compatible (antisymmetric) boundary conditions $G(\pm 7) = \pm 0.07$.  In this case there is a non-trivial solution for 
$c_1$ and $c_2$ and the solution relaxes to the corresponding steady state  (Figure \ref{fig:st}c).  

The next case is very instructive (Figure \ref{fig:st}d).  The initial condition is zero and symmetric negative boundary conditions are applied: $G(-7) = G(7) = -0.5$.  The steady solution compatible with the boundary conditions is shown as the dashed line.  It is obviously not reached as $t \to \infty$; instead negative vorticity grows exponentially.  Why?  In each half of the layer, $\eta < 0$ say, the steady solution has vorticity of both signs.  To reach the steady state, vorticity must change sign after it enters the left boundary and flows toward $\eta = 0$ (recall that we started with no vorticity in the domain).  However, neither advection, stretching, nor diffusion can change the sign of the vorticity.  
To reach the steady solution requires that the initial condition already contain the precise amount of vorticity of the opposite sign.  
To see this, consider Figure~\ref{fig:st}e where we started with the exact steady solution and applied one period of a sine wave as a perturbation:
\be
   G(\eta,t= 0) = G_2(\eta, a = 1.5) - 20\sin(2\pi\eta/\lambda), \hskip 0.5truecm \lambda = 2L.
\ee
The perturbation has zero area and so does not change the total vorticity.  As a result, the solution relaxes back to the steady solution.  On the other hand, in Figure~\ref{fig:st}f, the perturbation has non-zero area: it consists of three half-cosines:
\be
   G(\eta,t= 0) = G_2(\eta, a = 1.5) + \cos(2\pi\eta/\lambda), \hskip 0.5truecm \lambda = 2L/3,
\ee
and does not disturb the boundary conditions.  Note that the perturbation has more negative than positive area; the solution is unstable and negative vorticity grows exponentially.  It was verified that when the sign of the perturbation was changed, positive vorticity grew exponentially.  The results of the last two panels, (e) and (f), can be understood as follows.  Since the problem is linear, we may consider the evolution of the perturbation separately under homogeneous Dirichlet boundary conditions and that was already done in Figures \ref{fig:st}a and b.
There we saw that an initial condition with zero total vorticity decays while an initial condition with a non-zero total vorticity grows. 

\begin{acknowledgments}
I thank Prof. A. Migdal for sending me his manuscript and for his encouragement.  Drs. I. Kiviashvili and A. Wray performed the internal review for which I am grateful.
The data that support the findings of this study are available from the author upon reasonable request.
\end{acknowledgments}

\appendix*
\section{APPENDIX}

It is shown that, apart from one term, the equations satisfied by the velocity field in the local strain eigenframe are the Navier-Stokes equations and the average velocity obeys corresponding Reynolds-averaged equations.

Let us transform the Navier-Stokes equations (with density set to unity)
\ba
\frac{\p u_i}{\p t} + \frac{\p}{\p x_j}(u_i u_j)&=& -\frac{\p p}{\p x_i}, + \nu\frac{\p^2 u_i}{\p x_j^2}, \\
\frac{\p u_j}{\p x_j} &=& 0,
\ea
into strain eigenframe coordinates $\vec{\xi}$ and velocities $\vec{v}$ at a given Eulerian location $\vec{x}_0$, which is taken to be the origin without loss of generality.  The transformation is given by
\be
x_j = A_{jk}(t) \xi_k, \hskip 0.5truecm \mathrm{and} \hskip 0.5truecm u_j = A_{jk}(t) v_k,
\ee
where $A(t)$ is a rotation matrix which depends only on time for a given $\vec{x}_0$.
Using the fact that the transpose of a rotation matrix is its inverse, one finds that apart from the time derivative term, the rest of the terms in the Navier-Stokes equations retain their form in eigenframe coordinates:
\ba
v_k (A^{-1})_{ij} \dot{A}_{jk} + \frac{\p v_i}{\p t} + \frac{\p}{\p \xi_j}(v_i v_j)&=& -\frac{\p p}{\p \xi_i}, + \nu\frac{\p^2 v_i}{\p \xi_j^2},\\
\frac{\p v_j}{\p \xi_j} &=& 0. \eql{ns_se}
\ea
Ensemble averaging over a sample of points $\vec{x}_0$ and invoking the ergodic hypothesis that this average equals the time average at a fixed $\vec{x}_0$ in a statistically stationary flow gives, in the usual way, the time-independent Reynolds-averaged equation:
\ba
\overline{v_k (A^{-1})_{ij} \dot{A}_{jk}} + \frac{\p}{\p \xi_j}(\bv_i \bv_j) + \frac{\p R_{ij}}{\p x_j}&=& -\frac{\p \bp}{\p \xi_i}, + \nu\frac{\p^2 \bv_i}{\p \xi_j^2}, \eql{final}\\
\frac{\p \bv_j}{\p \xi_j} &=& 0,
\ea
where $R_{ij} \equiv \overline{\vp_i \vp_j}$ is the Reynolds stress.  It would be of interest to obtain the Reynolds stress term and the first term in \eqp{final} from the simulations and compare them with the mean terms.  Note that the matrix $A(t)$ can be chosen to be the rotation matrix for any desired frame of interest, not necessarily the strain eigenframe.

Lundgren~\cite{Lundgren_2008} considered the velocity difference $\vw(\vX,\vecr,t) \equiv \vu(\vX+\vecr,t) - \vu(\vX,t)$, where $\vX(t)$ is the position of a Lagrangian particle.  Equation (1) in Kolmogorov \cite{Kolmogorov_1941} shows that this is also what he had in mind.  Lundgren \cite{Lundgren_2008} showed that $\vw(\vX,\vecr,t)$ obeys the Navier-Stokes equation with respect to the separation vector $\vecr$:
\ba
\frac{\p w_i}{\p t} + \frac{\p}{\p r_j}(w_i w_j)&=& -\frac{\p p}{\p r_i}, + \nu\frac{\p^2 w_i}{\p r_j^2},  \eql{lns}\\
\frac{\p w_j}{\p r_j} &=& 0. \eql{li}
\ea
It may be useful to consider $\vw$ in the strain eigenframe, its statistics and scaling properties.  For that purpose, equation \eqp{ns_se} is still valid except that $\dot{A}(t)$ is the time derivative of the rotation matrix following a Lagrangian particle.

\clearpage
\bibliography{note.bib}
\end{document}